\documentclass[superscriptaddress, reprint, twocolumn%
reprint,
amsmath,amssymb,
aps,
prb,
]{revtex4-2}

\usepackage{graphicx}% Include figure files
\usepackage{dcolumn}% Align table columns on decimal point
\usepackage{bm}% bold math
\usepackage{siunitx}
\usepackage{placeins}

\begin{document}
	
	\preprint{APS/123-QED}
	
	%\title{Structural Signatures of an Anomalous Electronic Phase in 1\emph{T}-TaS$_{2}$}% Force line breaks with \\
    \title{Three-dimensional electronic domain correlations in 1\emph{T}-TaS$_{2}$}% Force line breaks with \\

    \author{Corinna Burri}
    \author{Henry G. Bell}
	\author{Faris Dizdarević}
	\author{Wenxiang Hu}
    \affiliation{PSI Center for Photon Science, Paul Scherrer Institute, 5232 Villigen PSI, Switzerland}%
	\affiliation{Laboratory for Solid State Physics and Quantum Center, ETH Zurich, 8093 Zurich, Switzerland}
    \author{Jan Ravnik}
    \author{Jakub~Vonka}
    %\author{Bill Pedrini} he decided that he prefers to be acknowledged
    \affiliation{PSI Center for Photon Science, Paul Scherrer Institute, 5232 Villigen PSI, Switzerland}%
    \author{Yasin Ekinci}
    \author{Shih-Wen Huang}
	\author{Simon Gerber}
	\email{simon.gerber@psi.ch}
    \affiliation{PSI Center for Photon Science, Paul Scherrer Institute, 5232 Villigen PSI, Switzerland}%
    \author{Nelson Hua}
	\email{nelson.hua@psi.ch}
	\affiliation{PSI Center for Photon Science, Paul Scherrer Institute, 5232 Villigen PSI, Switzerland}%
 	\affiliation{Institute for Quantum Electronics, ETH Zurich, 8093 Zurich, Switzerland}
	\date{\today}% It is always \today, today,
	
	\begin{abstract}{
    The interplay of nanoscale electronic domains underpins many emergent phenomena of quantum materials, including the competition between charge density waves (CDW) and superconductivity in high-$T_{\rm c}$ cuprates, or the storage of information in phase-change memory devices. 
    Coupling to electronic domains provides an observable for pinpointing key interactions, \textit{e.g.} affecting phase transitions. While the equilibrium phase diagram of 1\emph{T}-TaS$_{2}$---characterized by unique transport properties and varying degrees of CDW commensurability---has been studied extensively, an understanding of how the electronic domains in the bulk behave across phase boundaries is lacking. We reveal the three-dimensional evolution of electronic domains in \mbox{1\emph{T}-TaS$_{2}$} using temperature-dependent X-ray diffraction and reciprocal space mapping, complemented by structure factor simulations based on the Hendricks-Teller method. 
    With this methodology, we identify an increasing number of stacking faults near the phase transitions, and a growing fraction of dimerized layers in the commensurate phase upon cooling. We provide structural evidence that the CDW domains mediate the transport properties at phase boundaries, and that they also account for an anomalous intermediate electronic phase within the triclinic regime upon heating. As a paradigmatic material with potential in phase-change memory applications, our study underscores the importance of domain sizes and layer stacking in defining electronic behaviors of van der Waals materials.} 
	\end{abstract}
	
	\maketitle

\section{Introduction}
Understanding how charge-density wave (CDW) networks interact with the underlying atomic structure is crucial for unraveling the mechanisms behind phase transitions and for advancing precise material characterization. A ubiquitous technique for probing such nanoscale electronic and structural phenomena at surfaces is scanning tunneling microscopy (STM), that has been extensively used in condensed matter research~\cite{Pasztor2019, Wang2020_STM, Spera2020, Zhang2023, Hu2024}. A prominent example is the layered transition metal dichalcogenide 1\emph{T}-TaS$_{2}$, where STM serves as the primary method for imaging the characteristic polaron star CDW domains across its various electronic phases~\cite{Thomson1994, Ma2016, Gerasimenko2019, Butler2020, Geng2023}. For example, the nucleation of labyrinth domain walls from a commensurate insulating CDW state to a hidden metallic CDW phase can be directly visualized with STM~\cite{Ma2016, Gerasimenko_Nat_2019}. 
However, the principal limitation of STM is its inability to image what is beneath the surface layers. The interlayer stacking order of these phases, for instance, remains hidden as STM can only reveal the stacking of the truncation layer at the surface, leading to many open questions relating to whether the surface states are representative of the bulk structure~\cite{Burk1992,Perfetti2008,Stahl2020,Wang20202,Park2022,Park2023}.

\begin{figure}[t]
\includegraphics{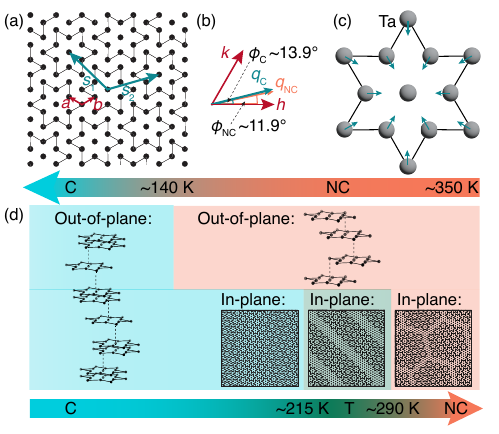}
\caption{(a) Schematic of the C state superlattice layer defined by Ta atom polaron stars, along with the lattice $a, b$ and superlattice $s_{1,2}$ vectors. 
(b) In-plane reciprocal lattice vectors $h,~k$, as well as CDW ordering vectors $q_\mathrm{C,NC}$ and their rotation angles $\phi_\mathrm{C,NC}$ with respect to $h$. 
(c) Sketch of a 13~Ta~atom polaron star with distortions toward the central atom. 
(d)~Evolution upon cooling and heating: the NC~phase above \SI{140}{\kelvin} shows spiral staircase out-of-plane order, whereas the C phase below features stacked dimers and monolayers. 
Upon heating, the intermediate T~phase emerges with in-plane stripe order.
}
\label{fig:Figure1a}
\end{figure}

We use X-ray diffraction (XRD) to complement STM, additionally revealing the correlation length along the third dimension, \textit{i.e.} the polaron stacking direction, through different CDW phases of 1\emph{T}-TaS$_{2}$. 
While standard XRD is incapable of directly imaging domains, it can obtain the average domain sizes in three-dimensions~(3D) due to its high penetration depth---fully illuminating exfoliated 1\emph{T}-TaS$_{2}$ flakes. 
In particular, we cool from the nearly-commensurate (NC) CDW phase at room temperature ($T = 325$~K) to the commensurate~(C) CDW state down to a base temperature of $T = 12$~K, before heating back to the NC state, traversing the triclinic (T) CDW phase along the way. 
This hysteretic sequence of CDW states is schematically shown in Fig.~\ref{fig:Figure1a}. 
The NC~phase is characterized by superlattice clusters of polaron stars that are separated by domain walls. The stacking arrangement of the stars can be classified into three types: When the central atoms of the stars are aligned directly on top of each other, the configuration is referred to as $T_\mathrm{A}$ stacking; if the central atom is positioned above the nearest-neighbor atoms of the star below, it is called $T_\mathrm{B}$~stacking; and when it is aligned above the next-nearest-neighbor atoms, the configuration is known as $T_\mathrm{C}$~stacking \cite{Butler2020, Lee2021, Hua2025}. 
In the NC phase, 13 Ta atoms form star-shaped domains where the superlattice ordering vector is rotated by $\phi_{\mathrm{NC}} = \SI{11.9}{\degree}$ \cite{Stahl2020} relative to the in-plane lattice vector. The the out-of-plane stacking follows a relatively ordered spiral staircase configuration~\cite{Hua2025} with a $T_\mathrm{C}$ stacking order \cite{Butler2020, Lee2021, Hua2025}, where the central Ta atom of a polaron star is stacked above one of the three symmetry-equivalent outer Ta atoms from below, as schematically depicted in Fig.~1(d). 
Around $140$~K, a first-order transition---evidenced by a step-like resistance increase and pronounced hysteresis in transport measurements~\cite{Wilson1975,Sipos2008}---leads to the insulating C~phase. There, the ordering vector is rotated by $\phi_{\mathrm{C}} = \SI{13.9}{\degree}$ \cite{Scruby1975, Stahl2020}, and the out-of-plane stacking consists of coexisting dimerized and monolayer domains~\cite{TOSATTI1976, Sipos2008, Lee2019, Hua2025}.

Upon heating, 1\emph{T}-TaS$_{2}$ undergoes a transition from the C phase to the T phase at $\approx$ \SI{215}{\kelvin} before relaxing back to the NC state around $\approx$ \SI{290}{\kelvin} \cite{Sezerman1980,Fung1980,Tanda1984,Nakanishi1984,Tanda1985}. 
In transport, the C-T transition is observed as a drop in both the in- and out-of-plane resistance at the end of the hysteresis loop~\cite{Wilson1975,Sipos2008}. 
The subsequent \mbox{T-NC}~transition shows another small step in the out-of-plane resistance~\cite{Manzke1989,Martino2020,Wang20202}, an out-of-plane lattice expansion~\cite{Sezerman1980} and a kink in the Seebeck coefficient~\cite{Tani1979}. 
In XRD, the T phase is characterized by three satellite peaks in the vicinity of each C peak and can be described as a stretched honeycomb lattice. STM measurements, on the other hand, show a faint striped phase~\cite{Burk1992,Thomson1994, Geng2023} on the surface, suggesting that the domain structure of the T phase is quite complex. Since the T phase appears only upon heating, it can be interpreted as a metastable state corresponding to a local instead of a global minimum in the material’s multivalley free-energy landscape~\cite{Tanda1984, Tanda1985, Nakatsugawa2019}.

Recently, an anomalous electronic phase of unknown origin, designated as the intermediate phase and appearing for  $T\approx215 - 235$~K, was discovered within the T~phase~\cite{Wang20202}. 
This was established via angle-resolved photoemission spectroscopy (ARPES) that probes the material's surface. 
Notably, the onset of the intermediate phase coincides with the resistance drop at the end of the hysteresis loop, presenting an intriguing contradiction: while ARPES reveals a band gap at the Fermi level, suggesting insulating behavior \cite{Wang20202}, transport measurements indicate that the material has already returned to the metallic resistance branch~\cite{Manzke1989}. 
Moreover, no distinct in- or out-of-plane features denoting the high-temperature end of the intermediate phase appear in transport, and no structural signatures from XRD were observed~ \cite{Wang20202}. 
The origin of the intermediate phase remains unclear as it was observed under different cleaving conditions, excluding a pure surface effect.

Here, we present temperature-dependent XRD measurements between $T = $ \si{12} and \SI{325}{\kelvin} of the C, NC, and T CDW phases upon cooling and heating. 
3D reciprocal space mapping (RSM) allows us to determine not only CDW peak positions and intensities, but also the associated correlation lengths that define the real-space domain sizes in three dimensions. 
In addition to standard peak fitting, we model the C and NC stacking by combining a Hendricks-Teller (HT) method  ~\cite{Hendricks1942,Treacy1991, Hua2025} with structure factor calculations to produce XRD patterns that match our experimental results. 

Our data reveals intensity signatures and anomalies in the CDW correlation lengths at the known first-order phase transitions, as well as the high-temperature end of the intermediate phase. 
This suggests that the anomalous intermediate phase within $T \approx 215 - 235$~K could originate from CDW domain interactions, illustrating how complex in- and out-of-plane stacking reconfigurations shape the rich phase diagram of 1\emph{T}-TaS$_{2}$.

\section{Experimental methods}

1$T$-TaS$_2$ bulk crystals are grown via chemical vapor transport with iodine as the transport agent ~\cite{Klanjsek2017}.
Single-crystal flakes are then mechanically exfoliated from bulk crystals using GelPak. 
Residual flakes are removed using Scotch tape to isolate a large, single flake. This single-crystal flake is subsequently transferred onto an oxidized silicon wafer (oxide thickness:~\SI{280}{\nano \meter}) and adheres through van der Waal forces (vdW). 
The studied flake has lateral dimensions of \mbox{56 $\times$ \SI{97}{\micro\meter\squared}} and a thickness of \SI{66}{\nano\meter}, measured using a profilometer, with the out-of-plane crystallographic direction perpendicular to the flake surface.

The XRD experiment is performed at the X04SA beamline at Swiss Light Source synchrotron of the Paul Scherrer Institute using a 5-circle diffractometer. 
The sample is cooled by a He flow cryostat connected to a hexapod for positioning with respect to the incident \mbox{X-ray} beam.
We use \SI{9.5}{\kilo\electronvolt} X-rays, avoiding fluorescence background from the Ta $L_3$ edge at 9.9 ~keV. Diffracton patterns are captured with  a 2D pixel detector (\emph{Eiger~S~500K}, \emph{DECTRIS}). The X-ray spot size is about \mbox{30 $\times$ \SI{100}{\micro\meter\squared}} that probes the entire flake at our incidence angle of \SI{5}{\degree}. 

Measurements of CDW peaks are performed upon cooling and heating between \si{12} and \SI{325}{\kelvin} with a \SI{1}{\kelvin / \minute} ramp rate.
At each temperature, rocking curve (sample rotation) measurements are conducted where the area detector captures the entirety of the CDW peaks. Rocking scans consist of 240 angular positions, uniformly distributed within a range of $2^\circ$ around each CDW peak, with an integration time of 1\,s per angular position.

The peak profiles are then reconstructed from 3D RSM by assigning the measured intensity (counts) of detector pixels with respective $(h k l)$ reciprocal space coordinates (see Supplementary Information of \cite{Burri2024} for details). 1D Lorentzian functions are subsequently fitted along the \textit{h}, \textit{k} and \textit{l} directions, from which the peak position, integrated intensity, and the full-width-at-half-maximum~(FWHM) are extracted with their respective errors. 
The correlation length, associated with the average CDW domain size, is determined by

\begin{equation}
\xi_{i} = \frac{a_i}{(FWHM_i)},
\label{Correlation_length}
\end{equation}

\noindent where $a_i$ is the respective lattice constant. 
Since we do not measure lattice reflections at every temperature, fixed values of $a=b=\SI{0.33}{\nano\meter}$ and $c=\SI{0.59}{\nano\meter}$~\cite{JELLINEK19629} are used for the entire measured temperature range. Several studies report less than \SI{0.2}{\percent} lattice changes upon cooling and heating~\cite{Givens1977,Sezerman1980,Guy1985,Wang2019}, which has a negligible impact on the RSM of the CDW satellite peaks in this study and do not account for the effects we observe. 

\section{Results and discussion}

\subsection{Temperature dependence of the CDW structure}

\begin{figure}[t]
\includegraphics{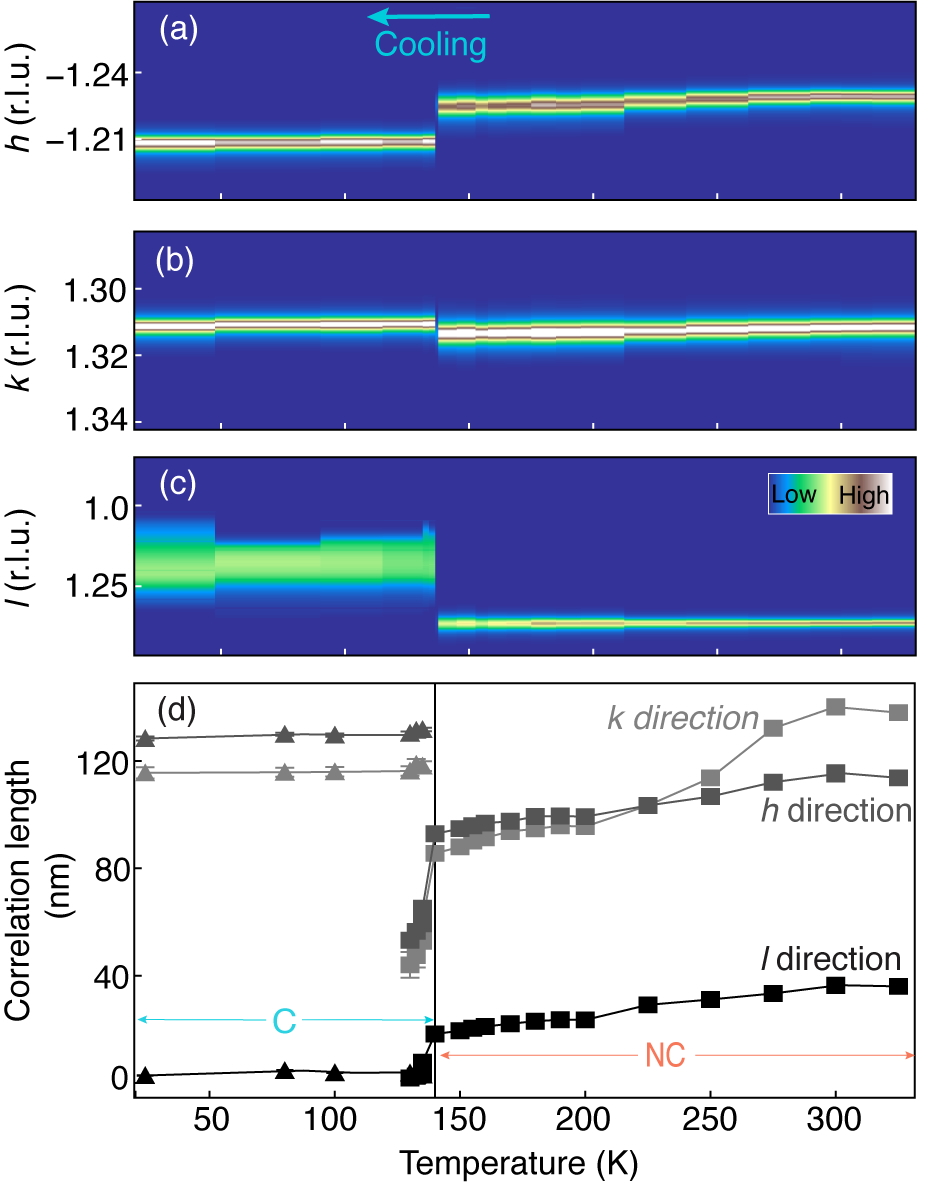}
\caption{(a)-(c) Projections of the NC and C$_\mathrm{I}$ CDW peaks along the reciprocal space directions upon cooling. 
Minimal and maximal intensities are 2 and 55~counts/s, respectively.
(d) Corresponding correlation lengths along the reciprocal space directions upon cooling. The solid vertical line denotes the metal-insulator CDW phase transition.}
\label{fig:Figure1}
\end{figure}

We track CDW peak centers about the $(\overline111)$ lattice peak, chosen to account for both in- and out-of-plane effects, as a function of temperature.
Figures \ref{fig:Figure1}(a)-(c) show the intensities of an NC and C reflection, that we denote as~C$_\mathrm{I}$, obtained by integrating a three-dimensional region of interest around each peak and projecting the resulting intensities onto the \textit{h}, \textit{k} and \textit{l} reciprocal space directions upon cooling. The NC peak position at \SI{300}{\kelvin} is determined from a fit to a Lorentzian, listed in Tab.~\ref{table:Table1}, that aligns with previous reports~\cite{Lauhle2015, Stahl2020}. 
Around \SI{140}{\kelvin}, the material goes through the first-order phase transition from the NC to the C~state observed as a discontinuous jump in the peak position. 
The C~state remains well-ordered along the in-plane \textit{h} and \textit{k} directions but exhibits broadening along $l$, associated with disordered out-of-plane stacking~\cite{Stahl2020, Hua2025}. 
The C$_\mathrm{I}$ peak position at \SI{100}{\kelvin} is listed in Tab. \ref{table:Table1} and agrees with the values in the literature~\cite{Lauhle2015, Stahl2020}.

\begin{figure}[t]
\includegraphics[width=\columnwidth]{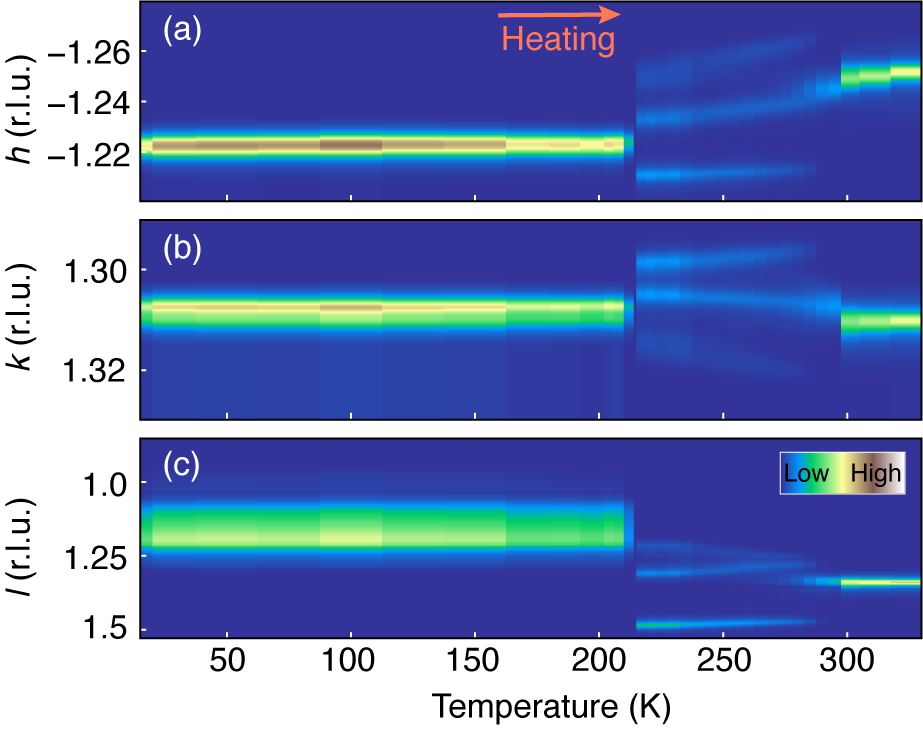}
\caption{(a)-(c) Projections of the C$_\mathrm{I}$, T$_\mathrm{I,II,III}$ and NC CDW peaks along the reciprocal space directions upon heating. 
Minimal and maximal intensities are 2 and 45~counts/s, respectively.}
\label{fig:Figure2}
\end{figure}

Upon heating [see Fig.~\ref{fig:Figure2}(a)-(c)], the C state remains stable up to \SI{215}{\kelvin}, where the transition to the T~phase occurs. 
At this point, three well-ordered T~peaks, designated as T$_\mathrm{I,II,III}$~\footnote{Considering that the different Ta~sites in the polaron stars of 1\emph{T}-TaS$_{2}$ are denoted as $T_{\rm A,B,C}$~\cite{Butler2020, Lee2021} and to avoid confusion with previous reports labeling the equilibrium CDW~phases with T$_{1,2,3}$~\cite{Scruby1975, POLLAK1976}, we choose the notation of the three T peaks as T$_\mathrm{I,II,III}$.}, emerge that have also been reported before \cite{Tanda1984,Wang20202} and are marked by a discontinuous change in peak position. The total intensity of the three T phase peaks is approximately the same as the C phase peak, indicating a conservation of scattering intensity across the transition. The positions are determined by fitting three Lorentzians and the results at \SI{220}{\kelvin} are listed in Tab.~\ref{table:Table1}. 
Upon heating, at around \SI{290}{\kelvin}, the material undergoes a further transition from the T to the NC~phase, as indicated by the appearance of the NC peak.

\begin{table}[t]
\centering
\begin{tabular}{|c||c|c|}
\hline \textbf{Peak} & \textbf{Peak position (r.l.u.)}\\
\hline\hline NC & $(-1.233, 1.312, 1.364)$\\
\hline C$_\mathrm{I}$ & $(-1.219, 1.311, 1.178)$ \\
\hline C$_\mathrm{II}$ & $(-1.299, 1.087, 0.841)$\\
\hline T$_\mathrm{I}$ & $(-1.225, 1.305, 1.485)$ \\
\hline T$_\mathrm{II}$ & $(-1.236, 1.298, 1.309)$ \\
\hline T$_\mathrm{III}$ & $(-1.245, 1.315, 1.218)$ \\
\hline
\end{tabular}
\caption{\mbox{(\textit{hkl})} coordinates in reciprocal lattice units~(r.l.u.) of the measured NC (\SI{300}{\kelvin}) and two C~peaks (\SI{100}{\kelvin}) upon cooling, as well as T phase peaks (\SI{220}{\kelvin}) upon heating. 
The positions are obtained by fitting Lorentzians to the data and yield uncertainties below $10^{-4}$~r.l.u.}
\label{table:Table1}
\end{table}

\subsection{Intra- and interlayer domains}

To extract the average electronic domain sizes across the different CDW states, we fit Lorentzians to the CDW peaks and determine the correlation lengths using Eq.~(\ref{Correlation_length}). 
Figure~\ref{fig:Figure1}(d) presents the correlation lengths along the \textit{h}, \textit{k} and \textit{l} directions during cooling. 
In the NC~state, the out-of-plane correlation length is shorter than the in-plane one, indicating that interlayer stacking of the polaron stars is more disordered than the intralayer domains. 
For example, the minimum of \SI{25}{\nano\meter} just above \SI{140}{\kelvin} along the $l$ direction corresponding to an average domain size that spans at least \SI{40} layers. 

Within a small temperature range of $10$~K below the first-order NC-C phase transition at \SI{140}{\kelvin}, the NC~domain size drastically decreases along all directions. 
The presence of a small, but measurable NC~peak indicates that some smaller NC domains are frozen within the C~phase. 
Upon further cooling, a discontinuous jump in the in-plane correlation lengths is observed, reaching a correlation length higher than that of the NC state before the transition. 
This is consistent with the disappearance of the domain wall network in the C~state~\cite{Gerasimenko2019}. 
However, surprisingly the overall magnitude of the in-plane ordering remains comparable to that of the NC~state at higher temperatures. 
Along the stacking direction, on the other hand, the correlation length shows a continuous drop, indicating high disorder in the C~state: 
The out-of-plane correlation length is $\approx$ $\SI{4}{\nano\meter}$, at least six times smaller than that of the NC~domains. 
Because of the commensurability of the C state, both the in- and out-of-plane correlation lengths remain nearly constant as a function of temperature. In contrast, the NC-state correlation lengths change noticeably upon cooling.

\begin{figure}[t]
\includegraphics[width=\columnwidth]{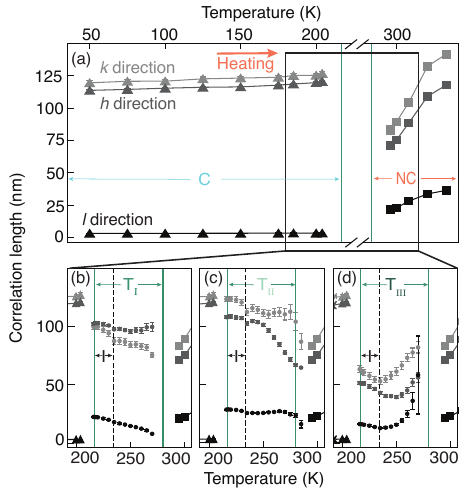}
\caption{In- and out-of-plane correlation lengths of the (a)~C,~NC, (b)~T$_\mathrm{I}$, (c)~T$_\mathrm{II}$, and (d)~T$_\mathrm{III}$ CDW peaks upon heating. Solid green lines denote the \mbox{C-T} and \mbox{T-NC} CDW transitions, whereas the vertical dashed line indicates the anomalous intermediate (I) to T~state transition.}
\label{fig:Figure3}
\end{figure}

Figure~\ref{fig:Figure3} shows the correlation lengths upon heating, displayed for C, NC and each of the T peaks. 
At the \mbox{C-T transition}, a discontinuity occurs along all directions. 
In general, the correlation lengths of the T~peaks fall between the corresponding values of the C~state just below the transition. 
In other words, the T phase domains appear more 3D than those of the C phase. 
Furthermore, unlike the correlation lengths in the C or NC~phases, the evolution of domain sizes in the T phase does not follow a monotonic trend, \textit{e.g.} those associated with the T$_\mathrm{III}$ peak [see Fig.~4(d)] show an inflection point. Interestingly, around \SI{235}{\kelvin}---within the T~phase---a dip of the in-plane correlation lengths tied to the T$_\mathrm{I,II}$ peaks is observed [see Fig.~4(b)-(c)]. 
This feature coincides with the high-temperature end of the previously reported intermediate phase~\cite{Wang20202}. 
As the \mbox{T-NC}~transition is approached, the T$_\mathrm{I,III}$ peaks disappear well before the T$_\mathrm{II}$~peak, preventing reliable fitting. 
But at the transition, the correlation lengths associated with the T$_\mathrm{II}$ peak seemingly follows a continuous crossover to the NC~state.

\begin{figure*}[t]
\includegraphics{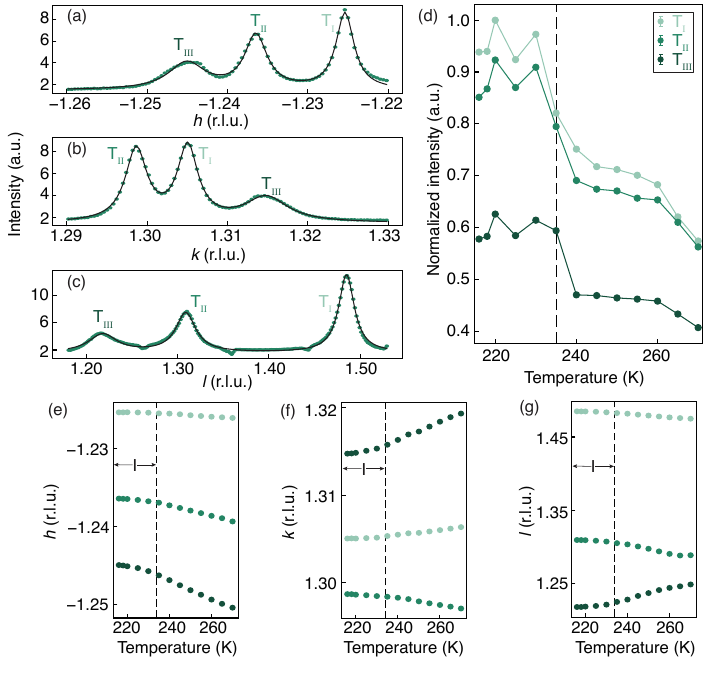}
\caption{(a)-(c) Lorentzian fits to the T$_\mathrm{I,II,III}$ peaks are shown at \SI{220}{\kelvin} along the \textit{h, k} and \textit{l} directions. (d) Normalized integrated intensity of the T~phase peaks upon heating, as well as (e)-(g) the peak positions in reciprocal space. The vertical dashed line at \SI{235}{\kelvin} denotes the high-temperature end of the intermediate phase.}
\label{fig:Figure4}
\end{figure*}

Figures~\ref{fig:Figure4}(a)-(c) show representative Lorentzian fits of the three peaks T$_\mathrm{I,II,III}$ at \SI{220}{\kelvin}, along with the temperature dependence of the integrated peak intensities [see Fig.~\ref{fig:Figure4}(d)], as well as the respective (\textit{hkl}) peak positions [see \mbox{Fig.~\ref{fig:Figure4}(e)-(g)}]. 
All peaks exhibit a smooth shift in position upon heating, with a more pronounced displacement along the out-of-plane direction, suggesting a stronger reordering of the stacking structure than in-plane. 
By examining solely the evolution of peak positions, as had  been done to find structural signatures of the intermediate phase~\cite{Wang20202}, there is no indication of distinct structural modifications. However, the changes in the peak intensities and the correlation lengths extracted from our 3D RSM reveal hints of the intermediate phase. 

Figure~\ref{fig:Figure4}(d) shows a drop in integrated intensity around \SI{235}{\kelvin} for all three T peaks, indicating a partial loss of CDW order, \textit{i.e.} that a fraction of the electronic domains disorders. This intensity change also coincides with the kink in the temperature dependence of the in-plane correlation lengths of peaks T$_\mathrm{I,II}$. 
While transport measurements do not show signatures of an intermediate phase, the onset of the T~phase is defined by a larger in-plane resistance change compared to the out-of-plane direction, and the \mbox{T-NC}~transition shows a drop in the in-plane resistance, but not out-of-plane~\cite{Martino2020}. 
Together, these observations suggest that the T~phase regime, which hosts the anomalous intermediate phase, is driven not only by restacking but also by in-plane effects. 
Also the intermediate phase seems to be defined by in-plane effects, albeit with more subtle origins tied to the changes in the electronic domain lengthscales. 
We note that the previous XRD study~\cite{Wang20202} focused on CDW peaks about the $(004)$ lattice peak that is predominantly sensitive to the out-of-plane direction whereas CDW peaks about $(\overline111)$ in this work contain in-plane components, allowing us to detect changes to the in-plane structure as well. 
Moreover, our findings suggest that the intermediate phase is not merely a surface phenomenon but extends throughout the bulk since our experiment probes the entire 66~nm thick flake.

\subsection{Structure factor calculations}

\begin{figure*}[t]
\includegraphics{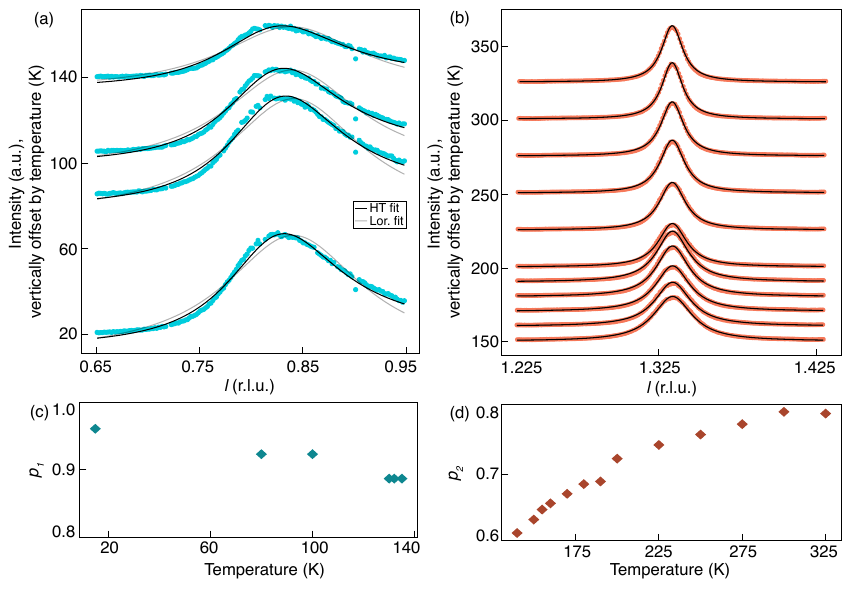}
\caption{HT results (black) and Lorentzian fits (grey) compared to the (a)~C$_\mathrm{I}$ and (b)~NC out-of-plane CDW projections measured upon cooling. (c) and~(d) show the extracted dimer fraction $p_1$ for the C~peak, and $T_C$~stacking probability $p_2$ for the NC~peak, respectively. We note that the error bars are smaller than the marker sizes.}
\label{fig:Figure3b}
\end{figure*}

To obtain further insights into the defining features of the domains, such as
the out-of-plane stacking order, we go beyond Lorentzian fitting and apply the HT~method~\cite{Hendricks1942,Treacy1991, Hua2025} to the NC and C~phases to construct probabilistic stacking sequences of \mbox{1$T$-TaS$_2$} planes.
We model a \mbox{1$T$-TaS$_2$} system consisting of 100~layers, corresponding to a thickness of $\approx$\SI{60}{\nano\meter}, which closely matches the sample measured in this study.
Calculating the structure factor of these systems then yields the XRD pattern that can be benchmarked against the experimental data. 
For the NC peak, we simulate a spiral staircase stacking of monolayers following the description in~\cite{Hua2025} and considering $T_\mathrm{C}$ stacking order, while varying the probability parameter $p_2$ (ranging from 0 to 1), where 0 implies no stacking faults and $1 - p_2$ represents the fraction of layers containing stacking faults that breaks the spiral staircase order.
Figures~6(a)-(b) show both the Lorentzian fits and the HT~result overlaid with our measurements of the C$_\mathrm{I}$ and NC peaks, clearly demonstrating that both approaches provide excellent agreement with the experimental data. 
The HT~result also gives a physical representation: $p_2 = 0.8$ at $T = 325$~K implies that, on average, the spiral staircase order is broken after every fourth layer ($\approx2$~nm).
This is a large discrepancy from the $\approx40$~nm correlation length obtained from the Eq.~(1), demonstrating that correlation lengths from peak fitting is informative on a average, but not an local scale. 
Figure~6(d) shows how the $p_2$~value decreases upon cooling, indicative of increasingly smaller domain sizes, that follows the same trend as the correlation lengths extracted from the Lorentzian fits in Fig.~2(d). The uncertainties are derived from the covariance matrix of the model fit and are approximately two orders of magnitude smaller than the corresponding $p_2$~values.

The C~phase, on the other hand, has been shown to follow a disordered $T_\mathrm{C}$~stacking~\cite{Stahl2020, Hua2025} with an additional complexity of featuring dimerized layers. 
We extract the dimer fraction, denoted as $p_1$, with the same variational approach where $p_1$ = 0.88 matches best with experimental data at $T = 140$~K [see Fig.~6(c)], \textit{i.e.} the majority of layers are dimerized while only about a tenth are monolayers. The uncertainties are again derived from the covariance matrix of the model fit and are four orders of magnitude smaller than the corresponding $p_1$ values. As the system is cooled further, the dimer fraction increases [see Fig. \ref{fig:Figure3b}(c)]. This finding aligns with earlier reports that used Monte Carlo simulations to show the number of dimerized layers depends on the cooling speed~\cite{Lee2019}. In our experiment, we indeed use a slow cooling and heating rate of \SI{1}{\kelvin / \minute}, and $p_1$ roughly agrees with the respective number of dimerized layers in~\cite{Lee2019}. 

Finally, the structure factor calculation of the C phase from the HT simulation matches the experimental data better than a Lorentzian fit, as illustrated in Fig.~\ref{fig:Figure3b}(b). 
We attribute the asymmetry of the peak to the tails of a secondary CDW peak at 0.5~r.l.u. that originates from the dimerized layers \cite{Stahl2020}. We note this secondary peak was not measured here and would have better constrained the HT~result. Also, a small fraction of $T_\mathrm{B}$~stacked layers have been found previously in \mbox{1$T$-TaS$_2$} flakes~\cite{Butler2020, Wang2023}, which is not considered here, but may contribute to the slight deviation between the experimental data and the HT~fit.

\section{Conclusions}
We systematically investigate the intra- and interlayer correlation lengths of electronic domains in \mbox{1\emph{T}-TaS$_{2}$} upon cooling and heating using X-ray based mapping of reciprocal space. Unsurprisingly, the CDW domain sizes remain quite constant within the C~phase, while with increasing temperature they increase in size in the NC~phase. 
Although the interlayer stacking is more ordered in the NC~phase compared to the C~phase, it is still much less correlated than the in-plane network of polaron stars. 
Respective correlation lengths are obtained by extracting the widths of Lorentzian fits. 
By modeling the XRD peaks using the HT~method and structure factor simulations, we obtain a physical basis of the domain structure  where the real-space stacking corresponds to $\approx$~80\% spiral staircase order in the NC~state at room temperature. 
Upon approaching the NC-C transition, this fraction decreases as more stacking faults are introduced. 
The C~state shows that $\approx$~90\% of layers are dimerized at the \mbox{NC-C~transition}, and that the dimer fraction increases further with cooling. 

We also identify electronic anomalies in CDW correlation lengths and peak intensities within the T~phase at \SI{235}{\kelvin} upon heating, which suggest that the anomalous, so-called intermediate phase may be driven by a complex reconfiguration of electronic domains. 
Specifically, we observe a dip in the in-plane correlation lengths, suggesting a reordering of the domain sizes, as well as a decrease in peak intensity of $\approx$~20\% which we speculate arises due to the material reaching a critical number of domain walls and hence a redistribution of the T domains. Next steps to solve the detailed nature of in-plane domain ordering and out-of-plane stacking in the intermediate and T~states, will involve extending the HT~approach with structure factor simulations to this phase. Here only three T~peaks were measured. However, to constrain both the in- and out-of-plane stacking which are both modified in the T~phase, it will be necessary to measure several T~peak triplets.

More generally, our study demonstrates that both in- and out-of-plane electronic order can play a central role in determining the phase diagrams of layered vdW materials. The respective intra- and interlayer domain sizes serve as sensitive probes of phase transitions, with peak fitting offering a basic estimate of correlation lengths. Complemented with physically-motivated models, such as the HT~approach, this provides a rigorous framework for interpreting stacking order. A systematic understanding of the nanoscale domain configuration of the equilibrium phases is essential not only to establish a microscopic description, but also to highlight controlled pathways through phase space towards desired non-equilibrium electronic states.

\section*{Acknowledgments}
We are grateful for fruitful discussions with G.~Aeppli, Y.~Cao and D.~Mihailovic. We acknowledge the Paul Scherrer Institute, Villigen, Switzerland, for providing synchrotron radiation beamtime at beamline X04SA of the Swiss Light Source. We thank P. Sutar for synthesizing the 1$T$-TaS$_2$ crystals, as well as B.~Pedrini for beamline and the PSI PICO operations team for technical support. This research was partially funded by the Swiss National Science Foundation~(SNSF) and the Slovenian Research And Innovation Agency (ARIS) as a part of the WEAVE framework under grant~213148 (ARIS project~\mbox{N1-0290}). J.R and N.H received funding from the European Union's Horizon 2020 research and innovation programme under the Marie Sklodowska-Curie grant agreements 701647~\mbox{(PSI-FELLOW-II-3i)} and 884104~(PSI-FELLOW-III-3i), respectively. C.B. and W.H. are grateful for funding from the European Research Council under Horizon 2020, within grant agreement 810451~(HERO).

%\clearpage

%\clearpage
\bibliography{Manuscript}% Produces the bibliography via BibTeX.

\end{document}